\documentclass[floatfix,
    reprint,
    superscriptaddress,
    amsmath,amssymb,
    aps,
]{revtex4-2}

\usepackage{graphicx}
\usepackage{dcolumn}
\usepackage{bm}
\usepackage{xcolor}
\usepackage{tikz}

\begin{document}

\preprint{APS/123-QED}

\title{Disordered origins, deterministic outcomes: \\ How the architecture of elastic networks imprints relaxed structure and mechanics}

\author{Stefanie Heyden}
\email{heydens@ethz.ch} 
\affiliation{Department of Civil, Environmental and Geomatic Engineering, ETH Zurich, 8092 Zurich, Switzerland}

\author{Mohit Pundir}
\affiliation{Department of Civil, Environmental and Geomatic Engineering, ETH Zurich, 8092 Zurich, Switzerland}

\author{Eric R. Dufresne}
\affiliation{Department of Physics, Cornell University, Ithaca NY 14850, USA}

\author{David S. Kammer}
\affiliation{Department of Civil, Environmental and Geomatic Engineering, ETH Zurich, 8092 Zurich, Switzerland}

\date{\today}

\begin{abstract}

This work targets the influence of disorder on the relaxed structure and macroscopic mechanical properties of elastic networks. 
We construct network classes of different types of disorder (length, topology and stiffness), which are subsequently equilibrated in a finite kinematics setting. 
Relaxed network structures are distinct among network classes, which opens the path towards exploiting easily accessible experimental measures as a way of inferring further microstructural details.

\end{abstract}

\maketitle

\section{Introduction}

Flexible networks are all around us: From polymers used in contact lenses \cite{Dung:2021}, to biomedical implants, soft robotics \cite{Minas:2021}, and gels making up our very own brain tissue \cite{Kuhl:2016}. 
At the core of accurately predicting material behavior and designing novel materials lies an accurate description of the material microstructure.
In addition to determining macroscopic mechanical properties, microstructural features are crucial in the realm of energetic stability \cite{Manning:2022}, force propagation within networks \cite{Ronceray:2016}, and eventually the onset of damage and fracture \cite{Mark:2003,Tehrani:2017}.
As such, microscopic details affect both the elastic as well as inelastic material response, with features such as a network's distribution of chain lengths eventually determining the onset of failure \cite{Tong:2018}.
In practice, the quantification of network structure is usually performed after network formation. This relaxed network state originates from a configuration initialized during polymerization, which may comprise different flavors of disorder (such as a network's length-, stiffness- and connectivity distribution). An enhanced understanding on how these defects affect relaxed network structure and macroscopic mechanical properties would greatly facilitate material characterization: it could allow to use easily accessible measures (such as local stiffness) to infer details on network structure that are more challenging to assess (such as the distribution in connectivity).

On the experimental side, the exact quantification of network structure poses a multi-lengthscale challenge, readily shown by the large spread of methods involved. 
At the scale of a few nanometers, multiple-quantum nuclear magnetic resonance and network disassembly spectrometry are established methods to distinguish local network structures such as dangling ends, loops, and ideal connections.
Common accessible measures at the macromolecular level include the distribution of chain lengths via double electron-electron resonance \cite{Drescher:2020} and radii of gyration using small angle neutron scattering \cite{Olsen:2023}.
Further indirect measures of network structure include the local elastic modulus (measured via, \emph{e.g.}, atomic force microscopy \cite{Millet:2021}), and network polydispersity (quantified via relaxometry of stressed networks or measurement of swelling pressure \cite{Bueche:1953,Gehman:1967}).

In the theoretical realm, early statistical works proposed an exponential form for the distribution of chain lengths in randomly crosslinked polymers \cite{Watson:1953,Watson:1954}.
These suggestions were supported by later studies on the equilibrium structure of randomly crosslinked networks via molecular dynamics simulations \cite{Kremer:1990}, as well as on the chain length distribution in multicomponent polymerization via Marcovian processes \cite{Tobita:1998}.
The gaussian approximation for the distribution of chain lengths was shown to hold for more complex coarse-grained polymer models taking into account the presence of entanglements \cite{Theodorou:2018}.
With regards to the effects of network topology, \cite{Zaccone:2016,Zaccone:2017} found that the local degree of inversion symmetry is a key player correlating strongly with the elastic response (besides connectivity and spring stretching).
Focusing on the resultant macroscopic material properties, analytical solutions for the elasticity of disordered central-force networks were derived in \cite{Zaccone:2011}.
Furthermore, polydisperse polymer networks were found to enhance ultimate mechanical network properties \cite{Mark:2003,Tehrani:2017}.
Knowledge of the exact chain length distribution was furthermore shown to be essential for correctly predicting the elastic properties \cite{Rovigatti:2023}, which readily highlights the importance of microstructural details on macroscopic network properties.
Within this realm, indirect measures such as a lognormal distribution of local elastic stiffness as found in biological tissues \cite{Millet:2021} immediately question their connection to network structure and disorder.

At the continuum scale, a large fraction of literature is dedicated towards incorporating microstructural information in continuum mechanics models.
In the simplest case, the affine model assumes chains as line segments being deformed according to the macroscopic deformation \cite{Treloar:1979}.
Averaging over all possible chain directions on the unit sphere was introduced as a means of breaking with affinity.
Different flavors of stretch averaging over the unit sphere range from varying measures of taking the average \cite{Puso:2003,Beatty:2003,Miehe:2004}, to taking into account different chain orientation distributions \cite{Rivlin:1948,Arruda:1993,Gasser:2005}.
In all cases, directional averaging depends on a relation between fiber orientation and deformation.
Recent works lift this dependence by computing the free energy of the continuum as an average of fibre free energy over the distribution of stretch \cite{Ehret:2022,Ehret:2023}.
Further advances investigate the effect of chain pre-stretch on structure-property relationships \cite{Zaccone:2019,Brassart:2024}.

However, an investigation in the reverse direction remains missing, i.e., the influence of microscopic network properties on relaxed network structure in equilibrium.  
This work aims to address the influence of disorder in flexible networks at the microscale on a generic level. 
Our analysis rests upon the assumption that systems are driven towards their energetic minimum.
To this end, we construct different network classes characterized by length-, stiffness- and topological disorder.
In all cases, we focus on the regime in which local connectivity does not fall below the rigidity threshold.
Networks are allowed to relax while obeying a macroscopic incompressibility constraint.
We find that the resultant network structure strongly differs among network classes.
This allows us to draw a direct relation between relaxed microstructure and initially induced disorder, opening the path towards exploiting easily accessible experimental measures as a means of inferring further microstructural details.

\section{Network Construction}

\begin{figure}
\begin{tikzpicture}
    \node[] at (0,2.9) {\color{darkgray}{I) length disorder}};
    \node (schematic) at (0,1.1) {\includegraphics[width=0.15\textwidth]{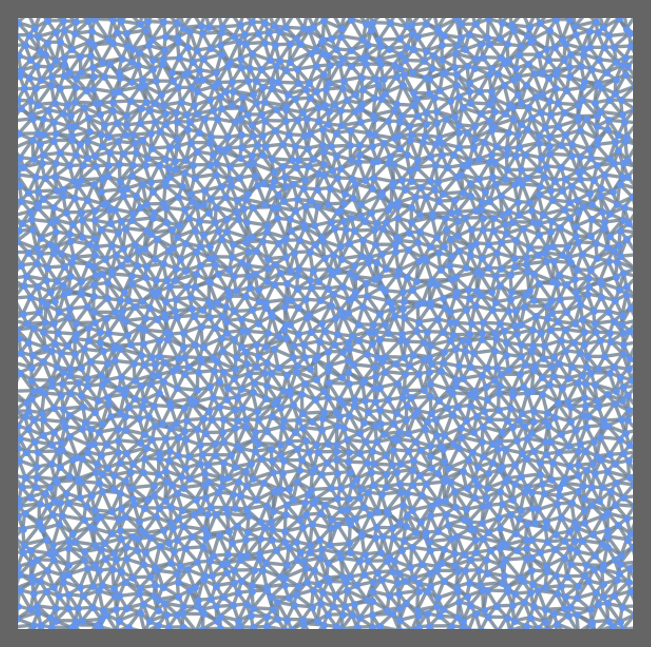}};
    \node[] at (3.5,3.3) {\color{darkgray}{II) length disorder}};
    \node[] at (3.5,2.9) {\color{darkgray}{stiffness disorder}};
    \node (schematic) at (3.5,1.1) {\includegraphics[width=0.15\textwidth]{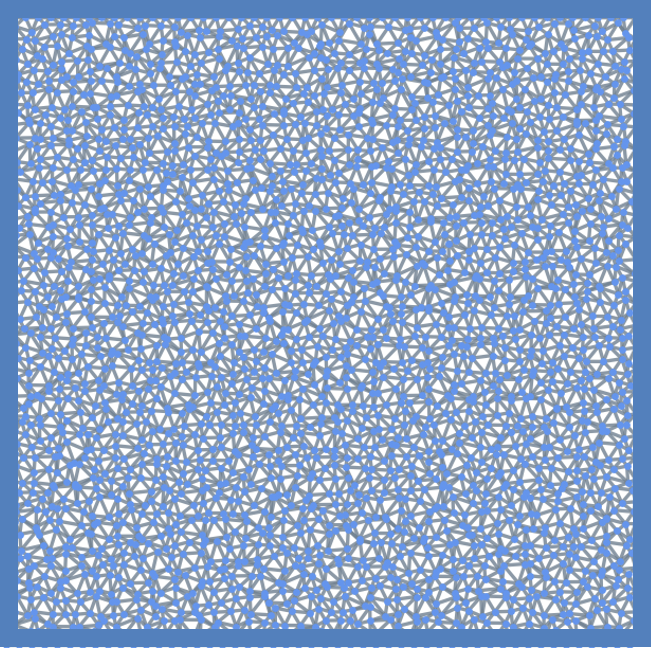}};
    \node[] at (0,-1.) {\color{darkgray}{III) length disorder}};
    \node[] at (0,-1.4) {\color{darkgray}{topological disorder}};
    \node (schematic) at (0,-3.2) {\includegraphics[width=0.15\textwidth]{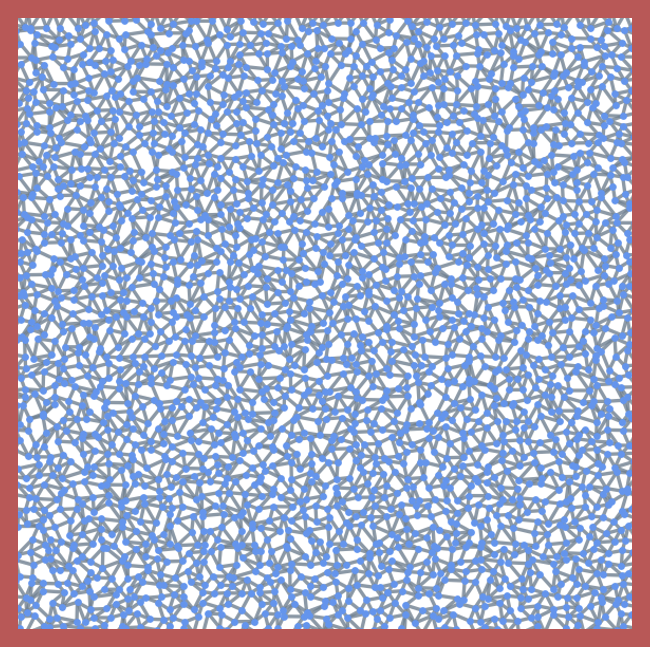}};
    \node[] at (3.5,-1.) {\color{darkgray}{IV) length disorder}};
    \node[] at (3.5,-1.4) {\color{darkgray}{topological disorder}};
    \node (schematic) at (3.5,-3.2) {\includegraphics[width=0.15\textwidth]{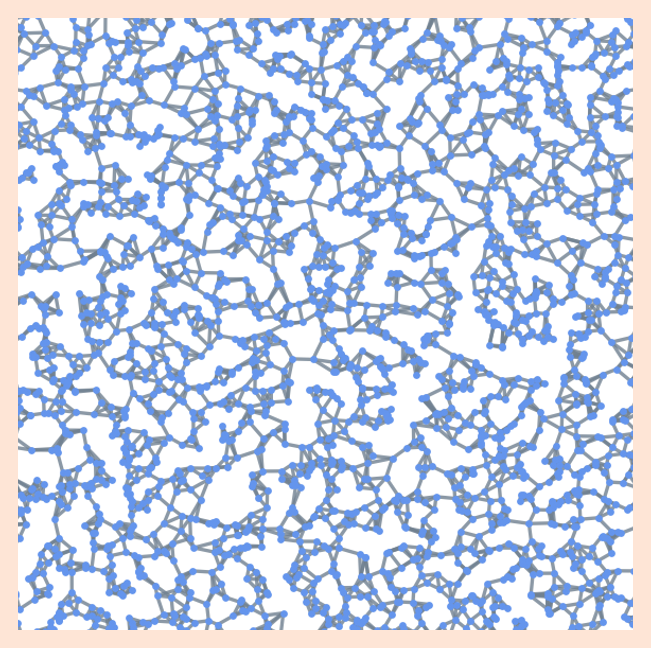}};
\end{tikzpicture}
\caption{Network classes under investigation (depicted in their initial configuration prior to relaxation).}
\label{fig:networks}
\end{figure}
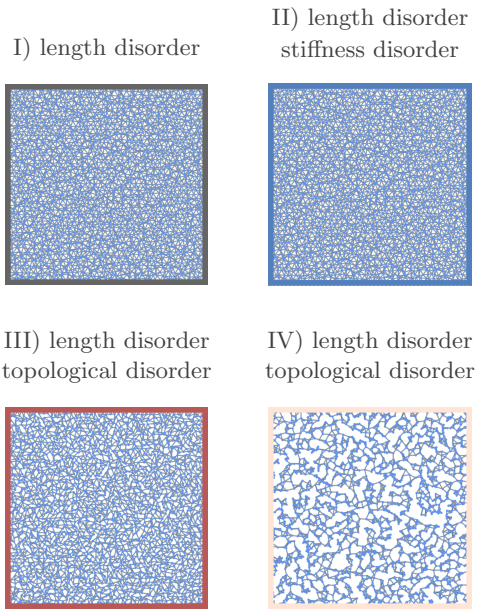

We construct network classes of varying disorder by two different network generation mechanisms: 
starting from regular triangular meshes with subsequently induced disorder, or via nearest neighbor searches connecting random nodes within a unit cell.
Each node within the network represents a crosslink and each edge constitutes an elastically active chain. 
In this context, the notion of spring length corresponds to the end-to-end distance of a molecular chain (which is not to be confused with the molecular weight).
The coarse-grained notion of spring stiffness is then defined by the molecular weight and persistence length of the molecular backbone.
For springs with stiffness $k^i$, undeformed length $l_0^i$ and deformed length $l_1^i$, elasticity is modeled via employing a (non)linear potential
\begin{align}
    \mathcal{E} &= \sum_{i=1}^N \int_0^{l_0^i} \frac{1}{2}k^i(\epsilon^i-\tilde{\epsilon})^2\mathcal{D}\,dl \nonumber \\
    &+ \left(\sum_{i=1}^N \int_0^{l_0^i} \frac{1}{4}k^i(\epsilon^i-\tilde{\epsilon})^4\mathcal{D}^3\,dl \right).
\label{eq:potential}
\end{align}
Here, $\epsilon^i=\frac{l_1^i-l_0^i}{l_0^i}$ is the strain upon deformation, whereas $\tilde{\epsilon}=\frac{\tilde{l}^i-l_0^i}{l_0^i}$ denotes the inherent eigenstrain for springs of restlength $\tilde{l}^i$.
Note that without any applied deformation ($l_1^i=l_0^i$), forces vanish for $l_0^i=\tilde{l}^i$.
For $\tilde{l}^i=0$ as taken in the subsequent analysis, each spring represents a coarse-grained freely jointed chain with vanishing bending rigidity and rest length zero \cite{Weiner:2017}.
It is important to note that by introducing a rest length zero, each length distribution within the network readily maps onto a strain distribution.
In addition, $\mathcal{D}$ is a switch function that allows to go from networks with pure length disorder to networks displaying stiffness disorder according to
\begin{align}
    \mathcal{D}=l_0^i \rightarrow \mathcal{E} &= \sum_{i=1}^N\frac{1}{2}k^i(l_1^i-\tilde{l}^i)^2 \nonumber \\
    &+ \left(\sum_{i=1}^N\frac{1}{4}k^i(l_1^i-\tilde{l}^i)^4 \right) \\
    \mathcal{D}=1 \rightarrow \mathcal{E} &= \sum_{i=1}^N\frac{1}{2}\frac{k^i}{l_0^i}(l_1^i-\tilde{l}^i)^2 \nonumber \\
    &+ \left( \sum_{i=1}^N\frac{1}{4}\frac{k^i}{l_0^3}(l_1^i-\tilde{l}^i)^4 \right).
\end{align}
In this way, four different disorder classes are introduced as illustrated in Figure~\ref{fig:networks}: 
Disorder class I features regular triangular networks with added nodal perturbations, inducing edge length disorder. 
The magnitude of nodal perturbation range $\delta x\in[-\delta l,\delta l]$, $\delta y\in[-\delta l,\delta l]$ is varied in relation to the unit cell size $l_0^{tri}$ of the regular triangular mesh.
Albeit not corresponding to polymeric materials, this highly idealized disorder class is used as a reference test case in the following.
Disorder class II displays the foregoing length disorder and adds further stiffness disorder. 
Polymer networks featuring controlled homogeneity in crosslinking (such as polyacrylamide or polyethylene glycol diacrylate) are prime examples of this disorder class.
Here, we add stiffness disorder as an inverse scaling with respect to edge length with $\tilde{k}^i=k^i/l_0^i$.
Disorder class III displays the previous length disorder and adds further topological disorder via randomly deleting a target percentage of edges, whereby the local connectivity is required to remain above the rigidity threshold.
Finally, disorder class IV introduces length- and topological disorder using a different network generation mechanism, namely a nearest neighbor search within a unit cell of randomly placed nodes. Here, each node is connected to its four closest neighbors, subsequently keeping the largest connected component as the main network and removing possible subgraphs.
Disorder classes III and IV correspond to polymeric materials displaying inhomogeneity in crosslinking distribution (such as polydimethylsiloxan), whereby strand polydispersity is maximized for the latter.

By way of construction, no dangling ends are present within the network, and chain elasticity is independent of loop order \cite{Rovigatti:2023}.
In the following, results are averaged over an ensemble of $10$ realizations of representative volume elements.
Further details on network statistics among different network classes are given in Table~\ref{tab:stats}.
Macroscopic incompressibility is enforced via fixing boundary nodes.
Finally, network relaxation is performed in a finite kinematics setting using \emph{Fenics} \cite{Fenics} with \emph{PETSc's} nonlinear equations solver \cite{petsc-efficient,petsc-user-ref,petsc-web-page}.

\onecolumngrid

\begin{figure}
\begin{tikzpicture}
    \node[] at (-10,-1.5) {\color{darkgray}{I) length disorder}};
    \node (schematic) at (-10,1) {\includegraphics[width=0.23\textwidth]{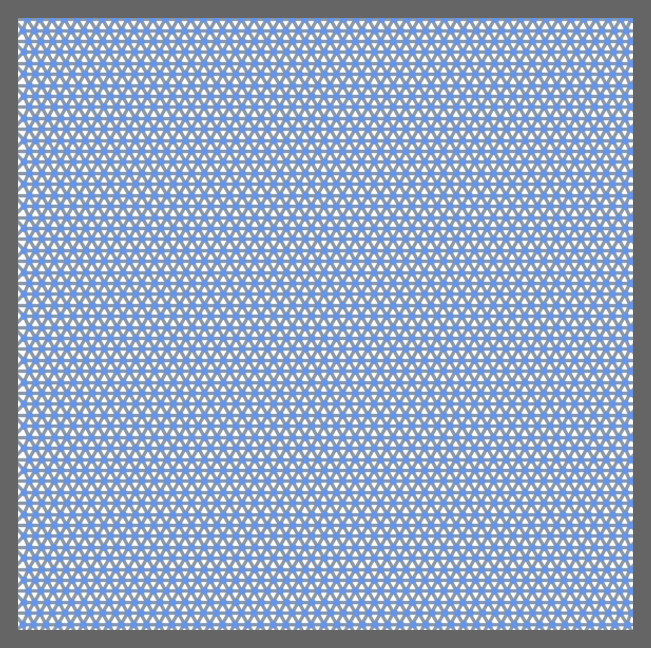}};
    \node[] at (-5.5,-1.5) {\color{darkgray}{II) length disorder}};
    \node[] at (-5.5,-1.9) {\color{darkgray}{stiffness disorder}};
    \node (schematic) at (-5.5,1) {\includegraphics[width=0.23\textwidth]{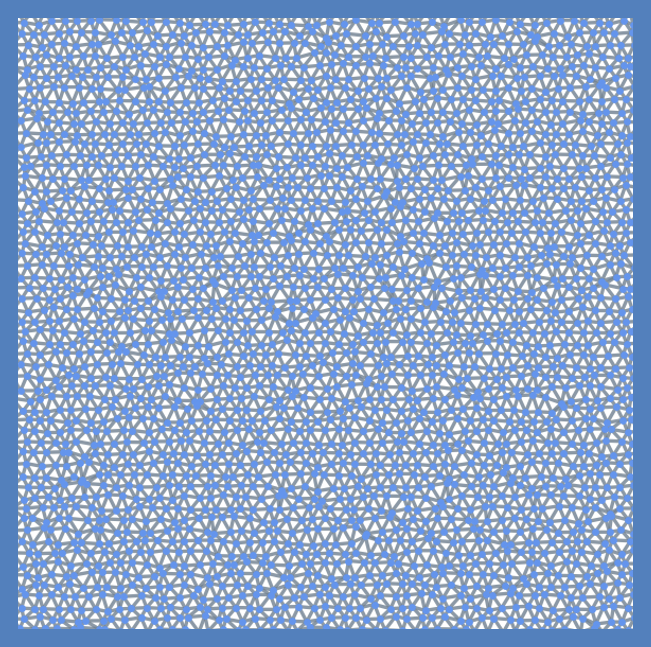}};
    \node[] at (-1,-1.5) {\color{darkgray}{III) length disorder}};
    \node[] at (-1,-1.9) {\color{darkgray}{topological disorder}};
    \node (schematic) at (-1,1) {\includegraphics[width=0.23\textwidth]{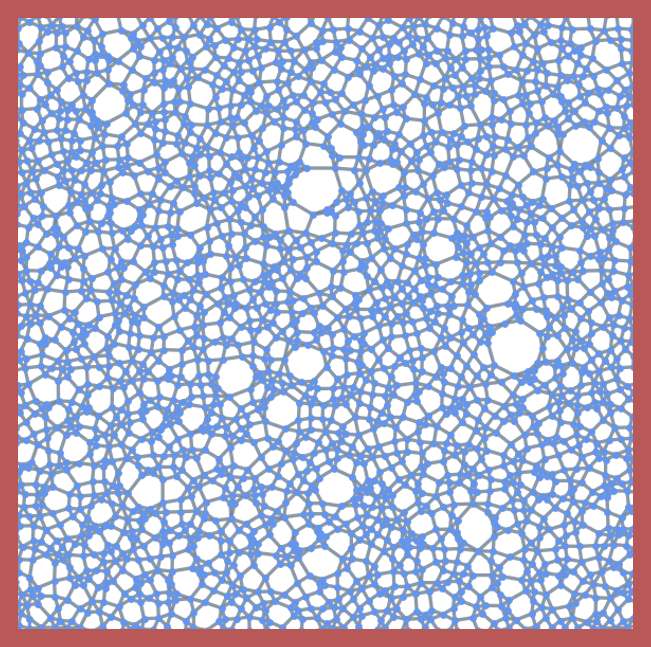}};
    \node[] at (3.5,-1.5) {\color{darkgray}{IV) length disorder}};
    \node[] at (3.5,-1.9) {\color{darkgray}{topological disorder}};
    \node (schematic) at (3.5,1) {\includegraphics[width=0.23\textwidth]{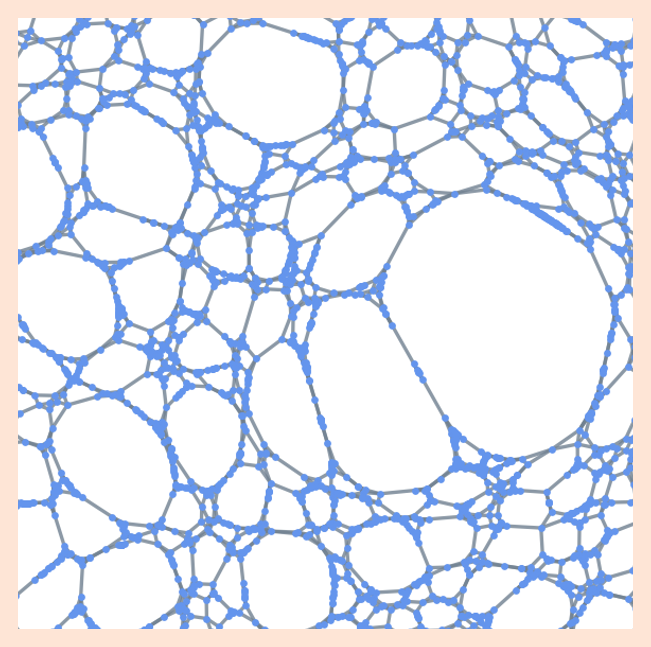}};
\end{tikzpicture}
\caption{Examplary plots of relaxed networks in their deformed configuration for different network classes under investigation.}
\label{fig:networks_deformed}
\end{figure}
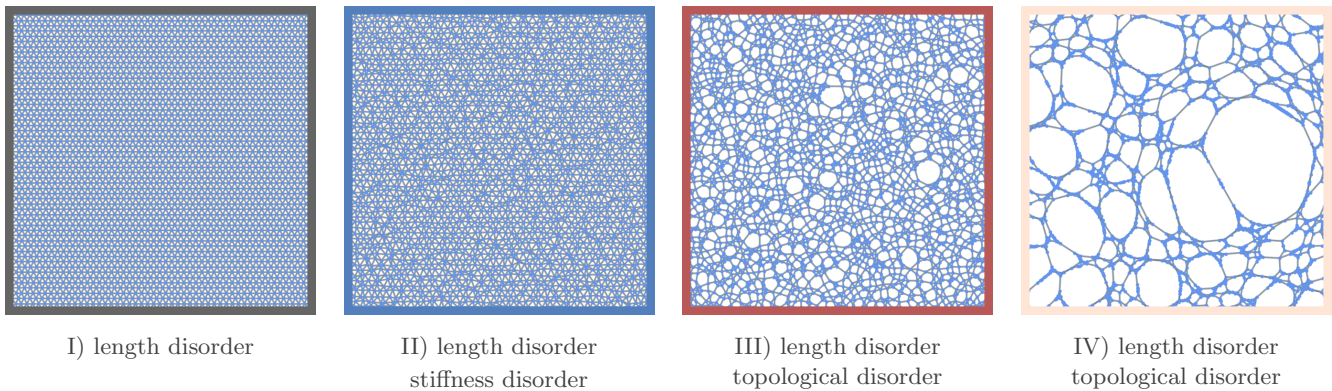

\twocolumngrid

\section{Results}

To investigate the influence of network disorder on micro- and macroscale network properties, we quantify relaxed length- and pore size distributions, as well as the emergent macroscopic shear modulus.

\subsection{Length distribution}

\begin{figure}
\begin{tikzpicture}
    \node (schematic) at (-1,7) {\includegraphics[width=0.5\textwidth]{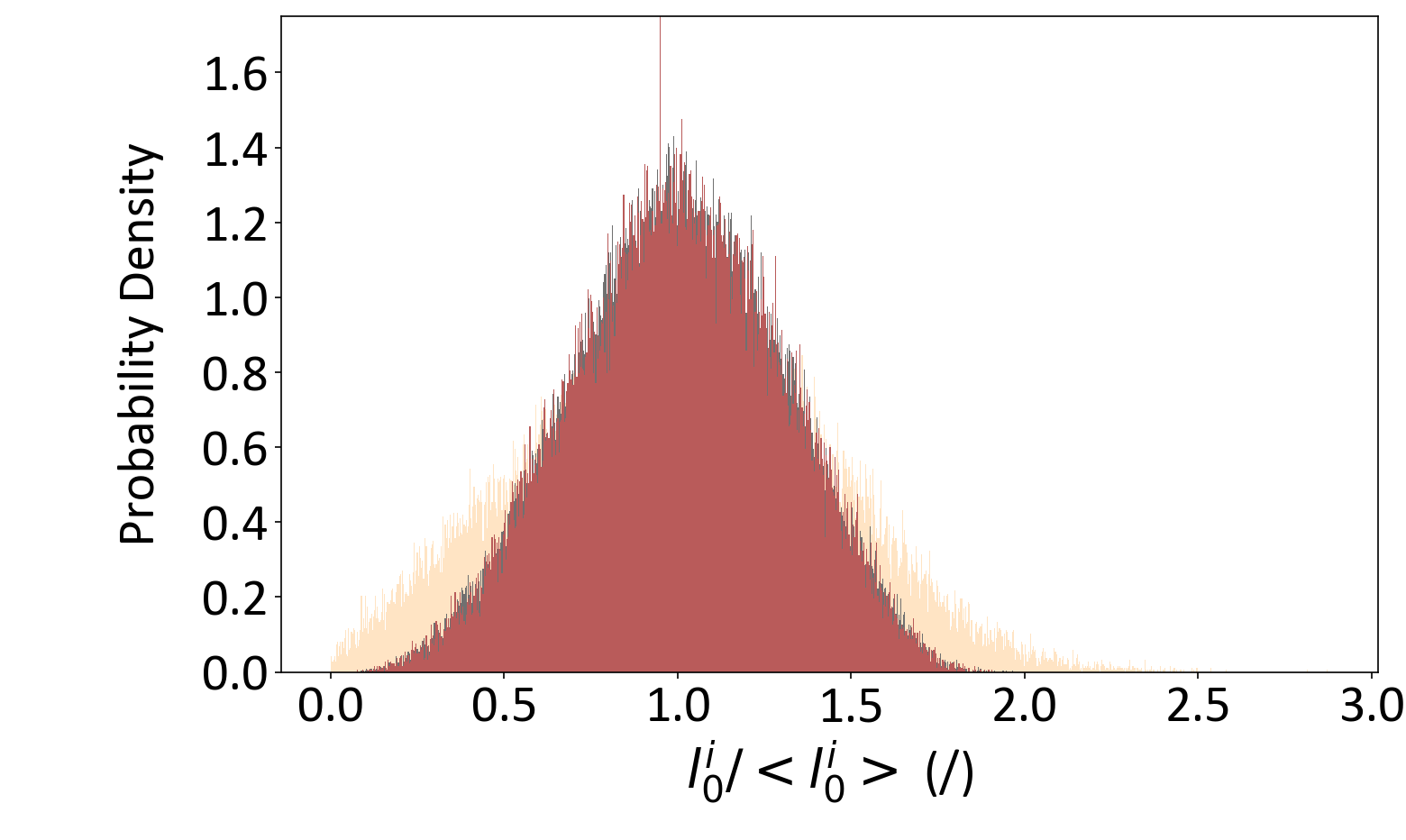}};
    \node (schematic) at (0,8) {\includegraphics[width=0.05\textwidth]{network_length_disorder_frame.png}};
    \node (schematic) at (1,8) {\includegraphics[width=0.05\textwidth]{network_stiffness_disorder_frame.png}};
    \node (schematic) at (-2.8,8.5) {\includegraphics[width=0.05\textwidth]{network_top_disorder_frame.png}};
    \node (schematic) at (1.5,6.1) {\includegraphics[width=0.05\textwidth]{network_top_disorder_2_frame.png}};
    \node (schematic) at (-1,1) {\includegraphics[width=0.46\textwidth]{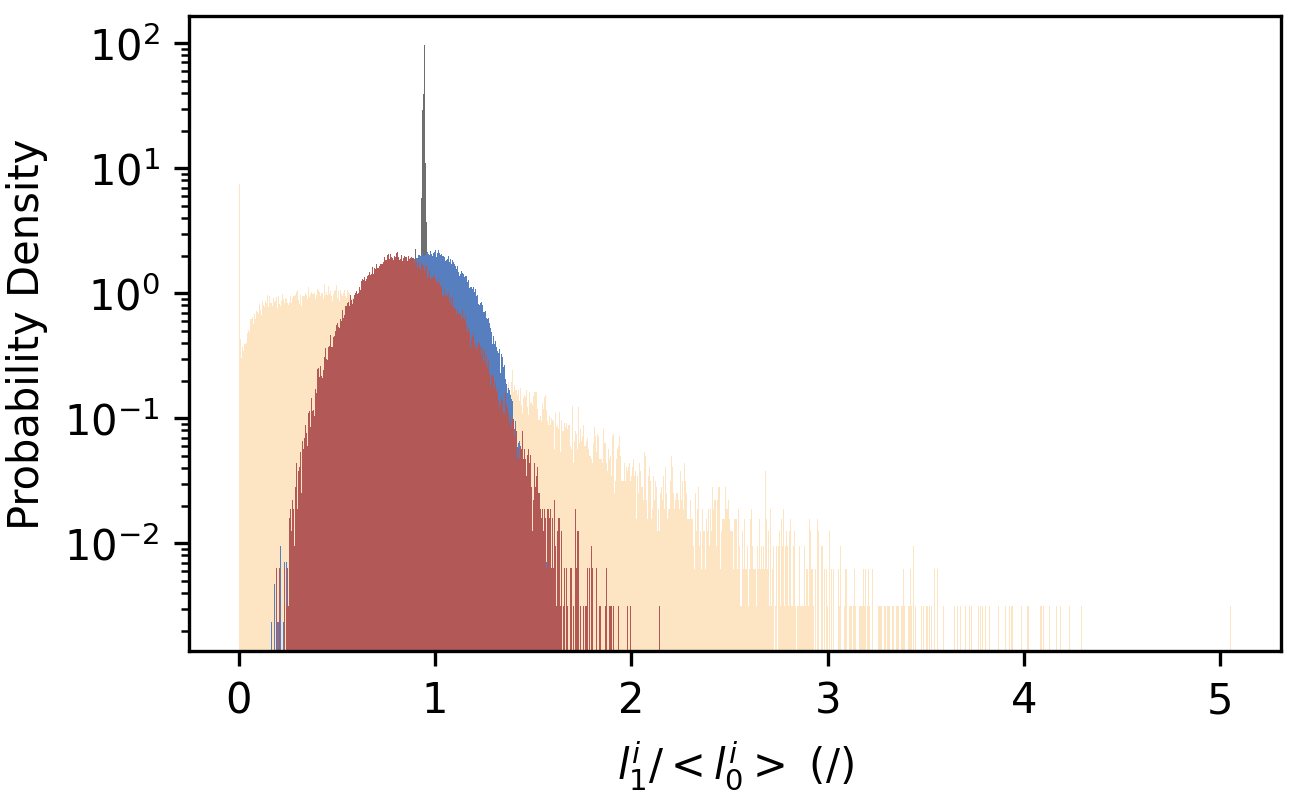}};
    \node (schematic) at (-1.8,2.8) {\includegraphics[width=0.05\textwidth]{network_length_disorder_frame.png}};
    \node (schematic) at (-1.3,1.5) {\includegraphics[width=0.05\textwidth]{network_stiffness_disorder_frame.png}};
    \node (schematic) at (-3.2,2.5) {\includegraphics[width=0.05\textwidth]{network_top_disorder_frame.png}};
    \node (schematic) at (1.5,0.5) {\includegraphics[width=0.05\textwidth]{network_top_disorder_2_frame.png}};
\end{tikzpicture}
\caption{Initial length distribution (top) and relaxed length distribution (bottom) for different network classes. Note that networks displaying length- and stiffness disorder (outlines in gray and blue, respectively) display the same initial distribution. Relaxed length distributions highlight a transition in relaxation characteristics (network of class II with $\delta l/l_0^{tri}=0.4$ and $100\%$ stiffness disorder,  network of class III with $25\,\%$ randomly deleted edges).}
\label{fig:histograms}
\end{figure}

Figure~\ref{fig:networks_deformed} shows examplary plots of relaxed networks in their deformed configuration for all different network classes under investigation.
In addition, Figure~\ref{fig:histograms} displays the corresponding initial and relaxed length histograms for the four network classes under consideration, employing a linear spring potential as given in Equation~\eqref{eq:potential} (the initial length disorder displays a gaussian length distribution across all network classes, \emph{cf.} Figure~\ref{fig:histograms}).
For networks of class I (pure length disorder induced on a regular triangular mesh), we expect relaxed configurations to return to the uniform distribution of an unperturbed triangular lattice.
Note that in this setting, the relaxed configuration is independent of spring stiffness $k^i$.
As highlighted in Figure~\ref{fig:histograms}, this behavior transitions to a gaussian length distribution for networks of class II, which also exhibit stiffness disorder. 
This is in direct relation to the way in which stiffness disorder is induced into the system, inversely scaling with edge length.
Interestingly, topologically disordered networks feature different relaxation characteristics, displaying lognormally- (network class III) and exponentially (network class IV) distributed edge lengths in their relaxed configuration.

\begin{figure}
\begin{tikzpicture}
    \node (schematic) at (0,1) {\includegraphics[width=0.5\textwidth]{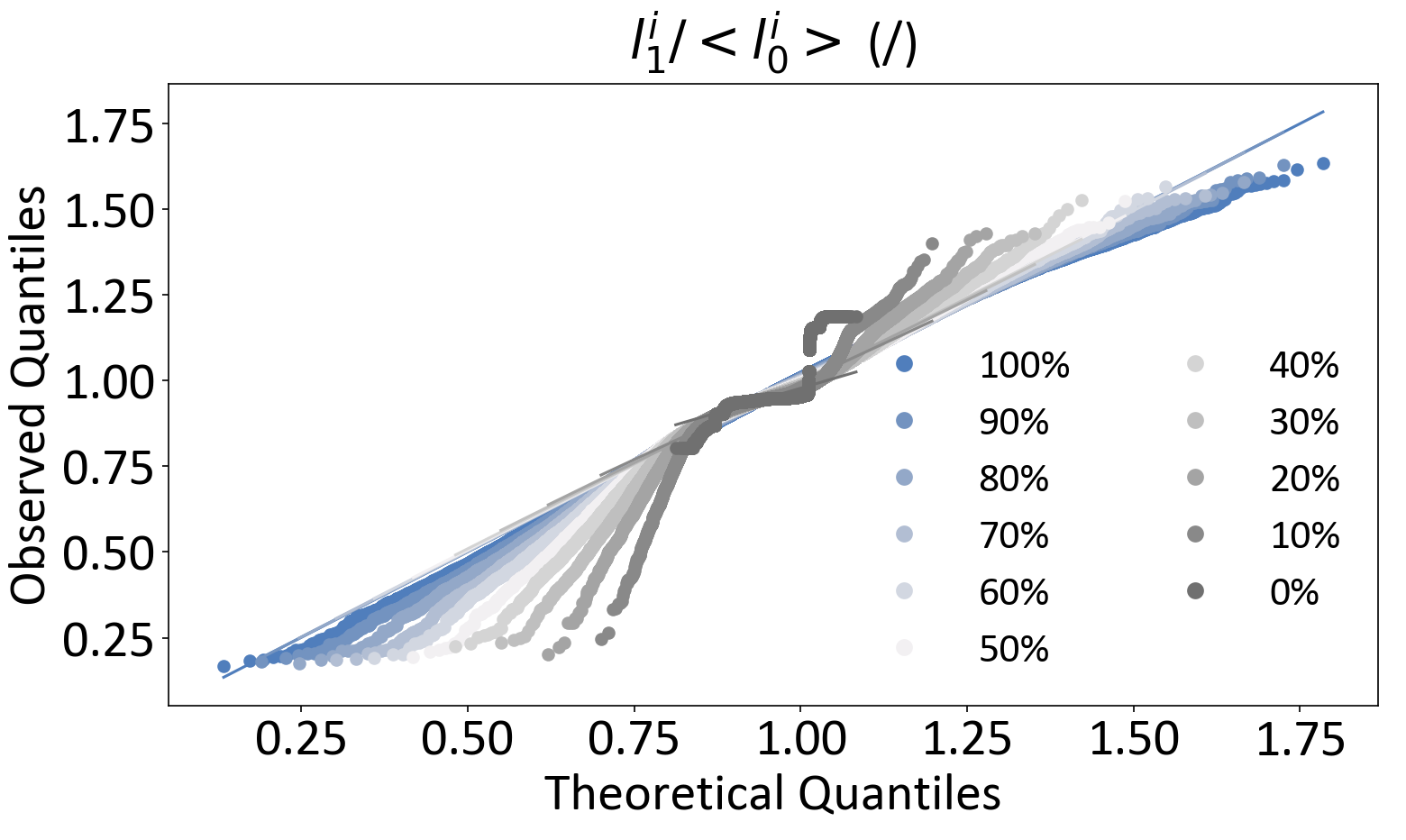}};
    \node (schematic) at (0.2,2) {\includegraphics[width=0.05\textwidth]{network_length_disorder_frame.png}};
    \node (schematic) at (3.8,1.9) {\includegraphics[width=0.05\textwidth]{network_stiffness_disorder_frame.png}};
    \node (schematic) at (0,-5) {\includegraphics[width=0.5\textwidth]{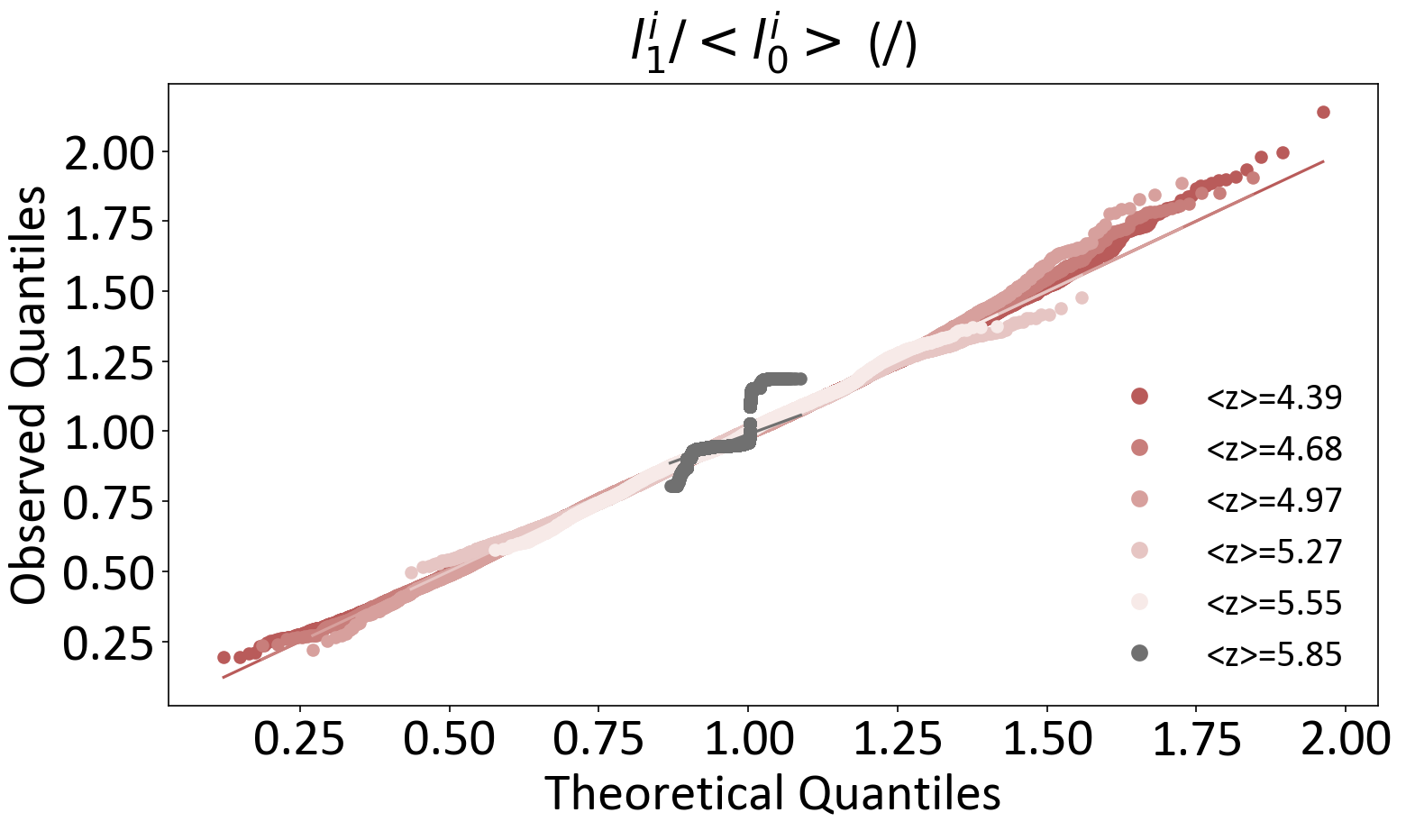}};
    \node (schematic) at (0.5,-4.1) {\includegraphics[width=0.05\textwidth]{network_length_disorder_frame.png}};
    \node (schematic) at (3.5,-4.2) {\includegraphics[width=0.05\textwidth]{network_top_disorder_frame.png}};
\end{tikzpicture}
\caption{Relaxed length distribution for networks of increasing stiffness disorder (top) and increasing topological disorder (bottom). Straight lines depict a comparison to normally distributed data (top) and lognormal distributed data (bottom).}
\label{fig:quantiles_length}
\end{figure}

\begin{figure}
\begin{tikzpicture}
    \node (schematic) at (0,1) {\includegraphics[width=0.5\textwidth]{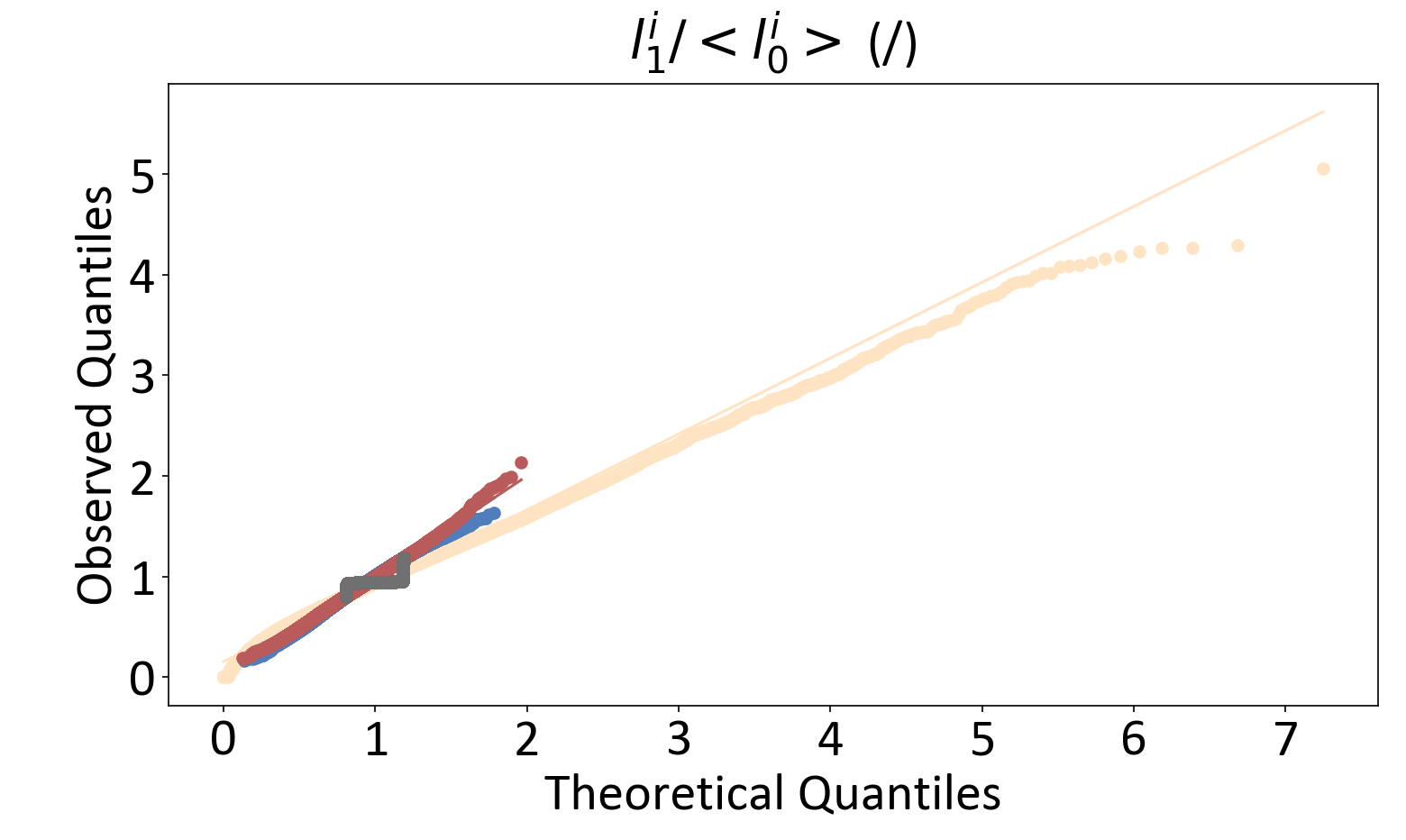}};
    \node (schematic) at (-2.5,0.7) {\includegraphics[width=0.05\textwidth]{network_top_disorder_frame.png}};
    \node (schematic) at (-0.8,-0.2) {\includegraphics[width=0.05\textwidth]{network_length_disorder_frame.png}};
    \node (schematic) at (-1,1.4) {\includegraphics[width=0.05\textwidth]{network_stiffness_disorder_frame.png}};
    \node (schematic) at (3.5,1.4) {\includegraphics[width=0.05\textwidth]{network_top_disorder_2_frame.png}};
    \node[] at (-2.5,1.4) {\small{\textit{lognormal}}};
    \node[] at (-1.,2.1) {\small{\textit{normal}}};
    \node[] at (0.3,-0.2) {\small{\textit{uniform}}};
    \node[] at (3.4,0.7) {\small{\textit{exponential}}};
\end{tikzpicture}
\caption{Relaxed length distribution of different network classes. Straight lines depict a comparison to distributions as noted for each case.}
\label{fig:quantiles_length_2}
\end{figure}

\begin{figure}
\begin{tikzpicture}
    \node (schematic) at (0,1) {\includegraphics[width=0.5\textwidth]{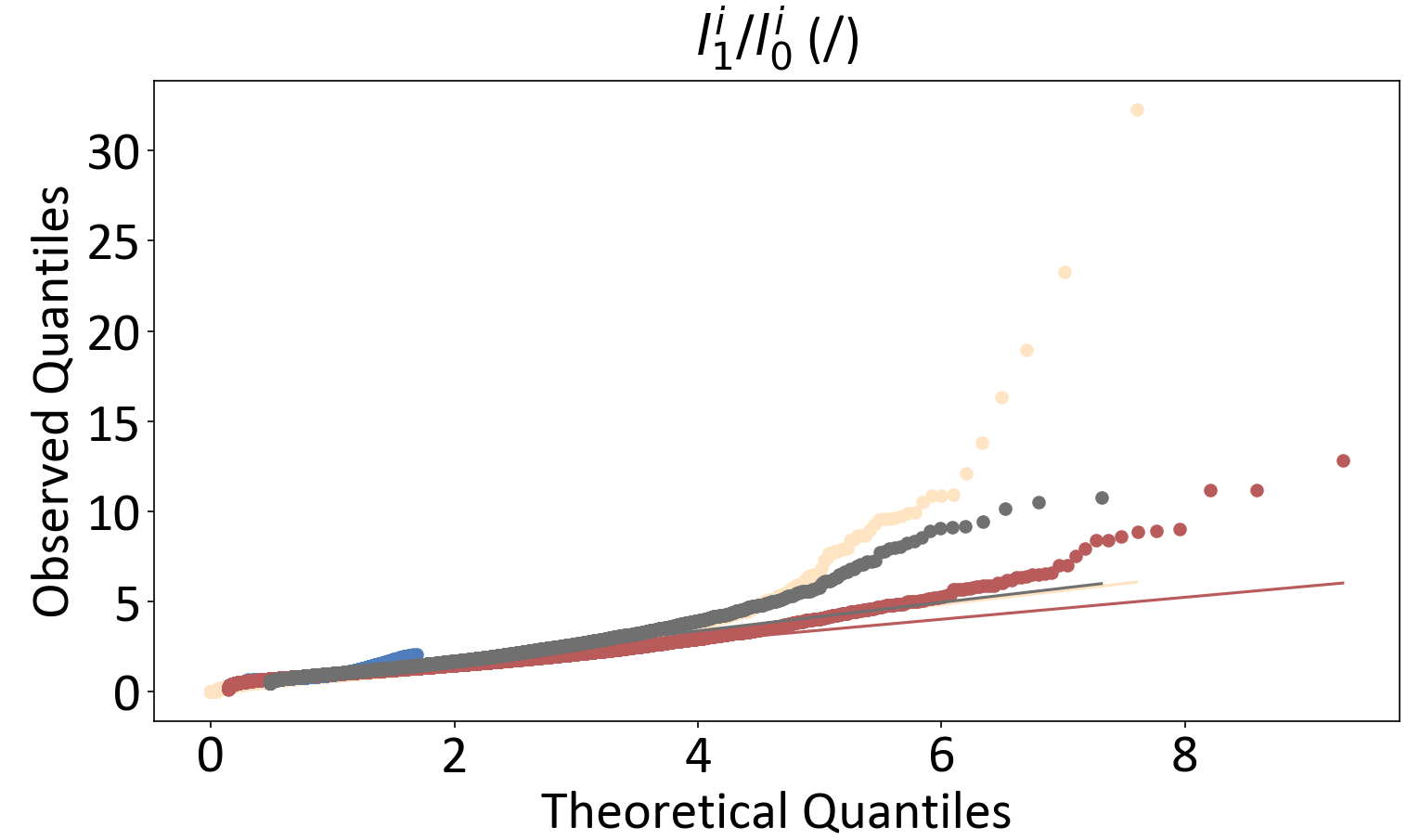}};
    \node (schematic) at (3.8,0.7) {\includegraphics[width=0.05\textwidth]{network_top_disorder_frame.png}};
    \node (schematic) at (-0.8,0.6) {\includegraphics[width=0.05\textwidth]{network_length_disorder_frame.png}};
    \node (schematic) at (-2.5,0.5) {\includegraphics[width=0.05\textwidth]{network_stiffness_disorder_frame.png}};
    \node (schematic) at (3.5,2.4) {\includegraphics[width=0.05\textwidth]{network_top_disorder_2_frame.png}};
    \node[] at (3.5,0.) {\small{\textit{lognormal}}};
    \node[] at (-2.5,-0.2) {\small{\textit{lognormal}}};
    \node[] at (-0.8,-0.1) {\small{\textit{lognormal}}};
    \node[] at (3.4,1.7) {\small{\textit{exponential}}};
\end{tikzpicture}
\caption{Relaxed stretch distribution of different network classes. Straight lines depict a comparison to distributions as noted for each case.}
\label{fig:quantiles_stretch}
\end{figure}

In order to quantitatively study the prevalent transition characteristics, we resort to quantile-quantile plots as shown in Figure~\ref{fig:quantiles_length}.
In these plots, the theoretical quantiles of normally distributed data (top) and lognormally distributed data (bottom) are tested against quantiles of the observed length distributions. 
We expect data to lie on a line if it is following the distribution tested against.
Figure~\ref{fig:quantiles_length} (top) displays relaxed length distributions of networks transitioning from class I (pure length disorder) to class II (length- and stiffness disorder), tested against the standard normal distribution.
As can be seen, the relaxed uniform length distribution ($0\,\%$ stiffness disorder) gradually shifts towards normally distributed lengths ($100\,\%$ stiffness disorder).
On the contrary, Figure~\ref{fig:quantiles_length} (bottom) displays relaxed length distributions of networks with varying topological disorder, tested against the standard lognormal distribution.
As shown, the relaxed uniform length distribution instantly shifts towards lognormally distributed lengths for networks displaying as little as $5\,\%$ topological disorder.
Further increasing the magnitude of topological disorder towards $25\,\%$ leads to a greater spread of data but is still highly aligned with the lognormal distribution.
Finally, networks of class IV displaying length- and topological disorder based on a nearest neighbor search during network generation relax to an exponential length distribution as illustrated in Figure~\ref{fig:quantiles_length_2}. 
This exponential relaxed length distribution displays tails leading up to much larger quantile ranges.
Note that stretch distributions within the sample remain to be exponential for networks of class IV, whereas all other network types display lognormal relaxed stretch distributions (\emph{cf.} Figure~\ref{fig:quantiles_stretch}).

The length distribution within samples has a direct relation to the microscopic stiffness distribution.
A lognormal distribution of elastic stiffness has been observed as a characteristic feature in biological tissues \cite{Millet:2021}, and furthermore been employed in higher level homogenization schemes \cite{Brassart:2024, Ehret:2022}.
Mathematically, logarithmic strains have been identified as the natural extension of strain measures to geometrical nonlinearities as encountered in the finite kinematics setting \cite{Neff:2016}.
Here, we find that topological disorder is the key driving force in departing from gaussian length distributions, and that the prevalent shift is induced instantly even at low levels of topological disorder.

\subsection{Macroscopic mechanical properties}

To further investigate the influence of relaxed edge length distribution on macroscopic mechanical properties, we perform simple shear tests on relaxed representative volume elements. 
The resultant shear modulus $\mu$ is extracted from the standard relation $<\boldsymbol{\sigma}>=\mu<\boldsymbol{\epsilon}>$, where the average stress of representative volume elements is computed as
\begin{equation}
    <\boldsymbol{\sigma}> = \frac{1}{\sum_{i=1}^Nl_1^i}\int_0^{l_1^i}\sum_{i=1}^N\sigma^i\boldsymbol{L}_1^i\otimes\boldsymbol{L}_1^i.
\end{equation}
Here, $l_1^i$ is the relaxed edge length, $\boldsymbol{L}_1^i$ denotes the relaxed edge tangent, and $\sigma^i$ is the stress within edges upon simple shear deformation.

\begin{figure}
\begin{tikzpicture}
    \node (schematic) at (0,0) {\includegraphics[width=0.5\textwidth]{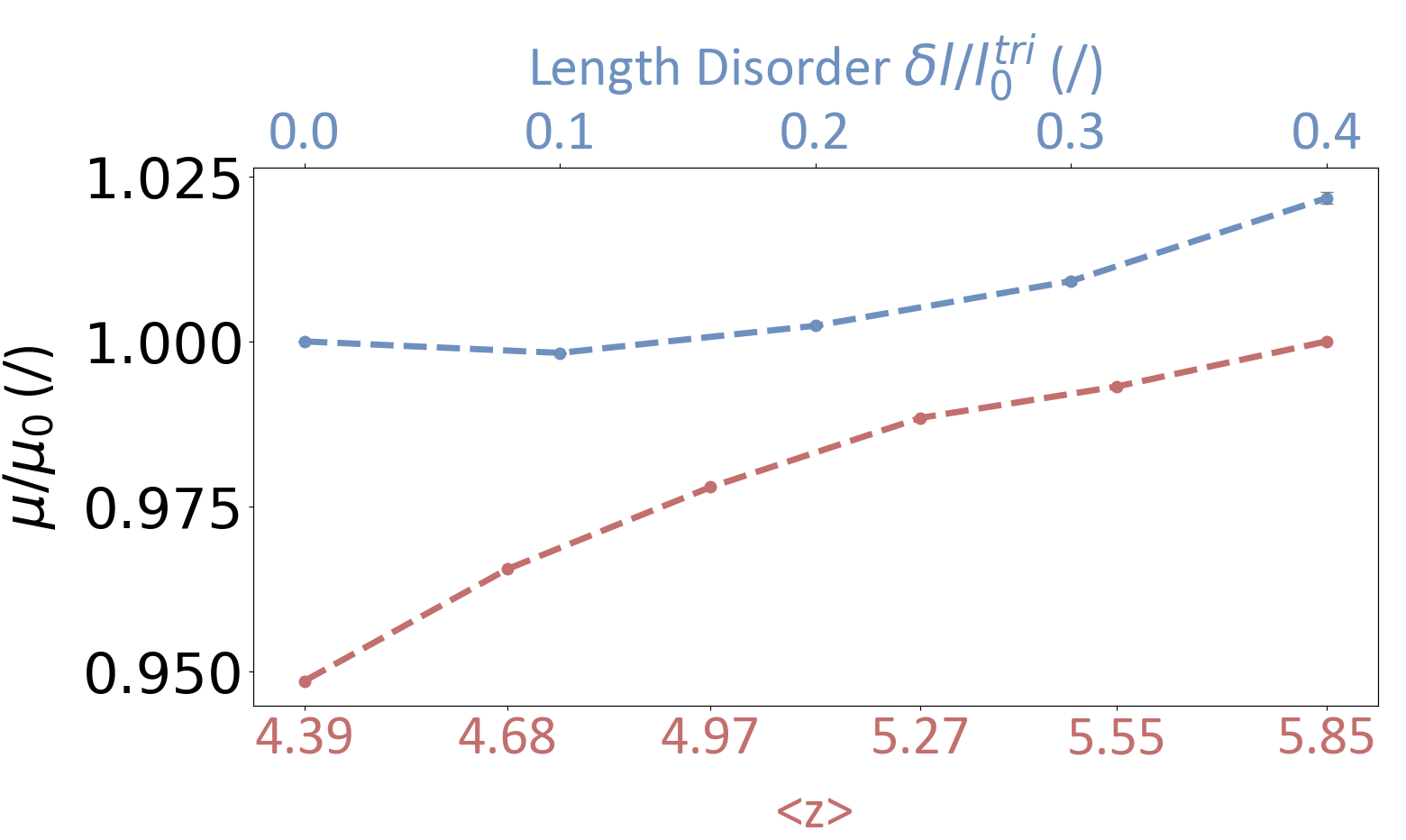}};
    \node (schematic) at (3,-0.5) {\includegraphics[width=0.05\textwidth]{network_top_disorder_frame.png}};
    \node (schematic) at (-2.48,0.5) {\includegraphics[width=0.01\textwidth]{network_length_disorder_frame.png}};
    \node (schematic) at (4,0.5) {\includegraphics[width=0.01\textwidth]{network_length_disorder_frame.png}};
    \node (schematic) at (-2,-0.2) {\includegraphics[width=0.05\textwidth]{network_stiffness_disorder_frame.png}};
\end{tikzpicture}
\caption{Influence of length disorder (top) and topological disorder (bottom) on macroscopic shear modulus. Increasing length disorder is induced in networks displaying $100\%$ stiffness disorder. Length disorder is varied by increasing the magnitude of nodal perturbation range $\delta x\in[-\delta l,\delta l]$, $\delta y\in[-\delta l,\delta l]$ in relation to the unit cell size $l_0$ of the regular triangular mesh.
Results are normalized by the shear modulus $\mu_0$ of a regular triangular mesh of linear elastic springs \cite{Alzebdeh:1999}.}
\label{fig:stiffness}
\end{figure}

Figure~\ref{fig:stiffness} shows the relaxed shear modulus for topologically disordered networks, as well as for fully stiffness disordered networks with increasing magnitude in length disorder. 
We find that inducing topological disorder reduces the relaxed shear modulus, whereas increasing length disorder within the system increases the relaxed macroscopic shear modulus. 
This is in contrast to investigations on length disorder in periodic trusses, which was found to decrease the continuum shear stiffness \cite{Glaesener:2023}.
In their analysis, \cite{Glaesener:2023} use beams to model individual truss members, which are assumed to be stress-free in their reference configuration. 
The inherent eigenstrain of network members in the present analysis might hence play an important role in macroscopic stiffness.
In comparison, topologically disordered networks based on a nearest neighbor search (network class IV) display a much lower macroscopic stiffness of $\mu/\mu_0\sim0.744$.

\subsection{Pore size distribution}

\begin{figure}
\begin{tikzpicture}
    \node (schematic) at (0,0) {\includegraphics[width=0.5\textwidth]{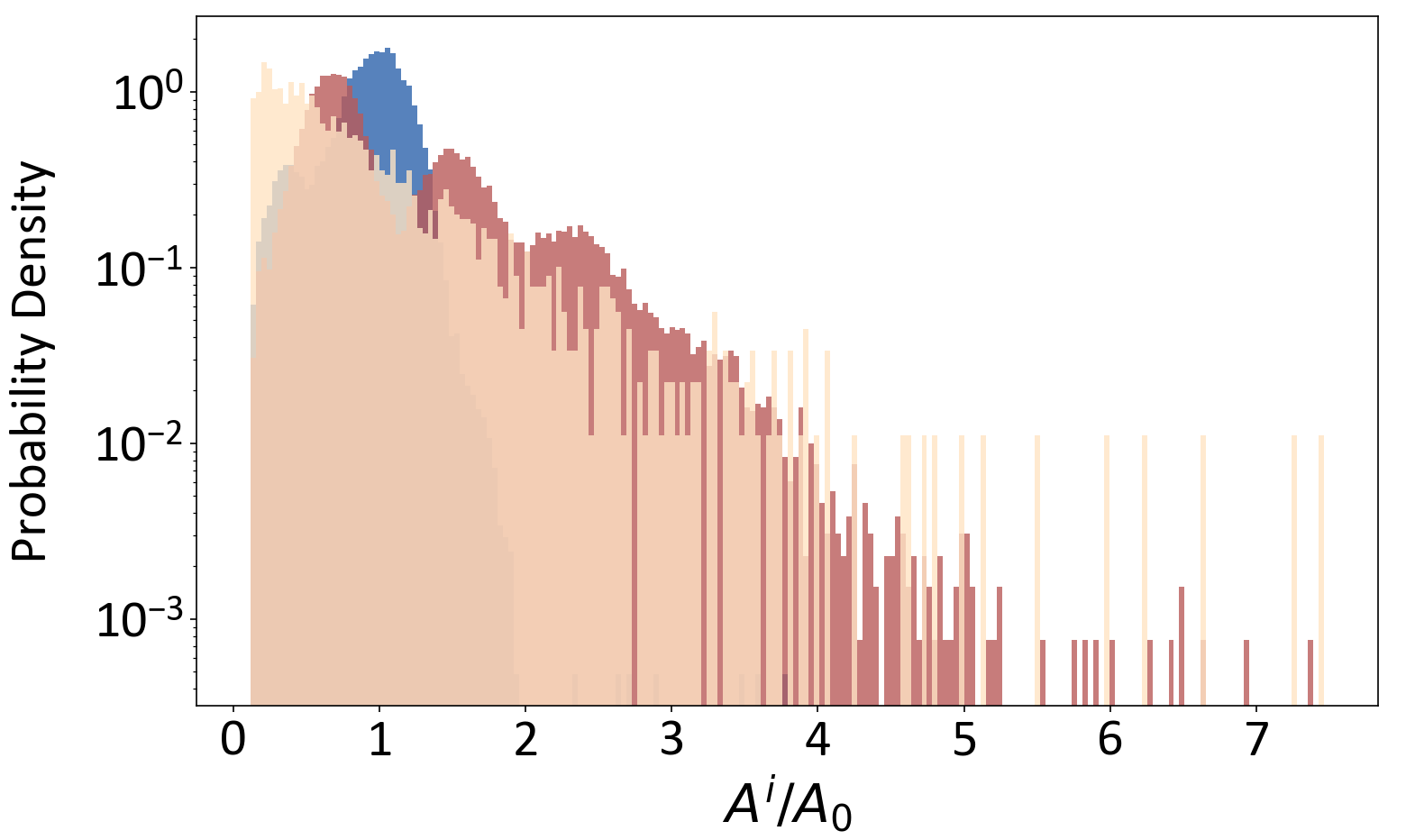}};
    \node (schematic) at (-0.7,1.9) {\includegraphics[width=0.06\textwidth]{network_stiffness_disorder_frame.png}};
    \node (schematic) at (2.5,-0.8) {\includegraphics[width=0.06\textwidth]{network_top_disorder_frame.png}};
    \node (schematic) at (1.4,1.) {\includegraphics[width=0.06\textwidth]{network_top_disorder_2_frame.png}};
\end{tikzpicture}
\caption{Relaxed pore size distribution for networks displaying stiffness disorder (disorder class II, $100\%$ disorder), topological disorder (disorder class III, $25\%$ deleted edges) and topological disorder induced by a nearest neighbor search (disorder class IV)). Pore sizes are normalized by the cell size $A_0$ within a regular triangular network.}
\label{fig:pores}
\end{figure}

\begin{figure}
\begin{tikzpicture}
    \node (schematic) at (0,1) {\includegraphics[width=0.5\textwidth]{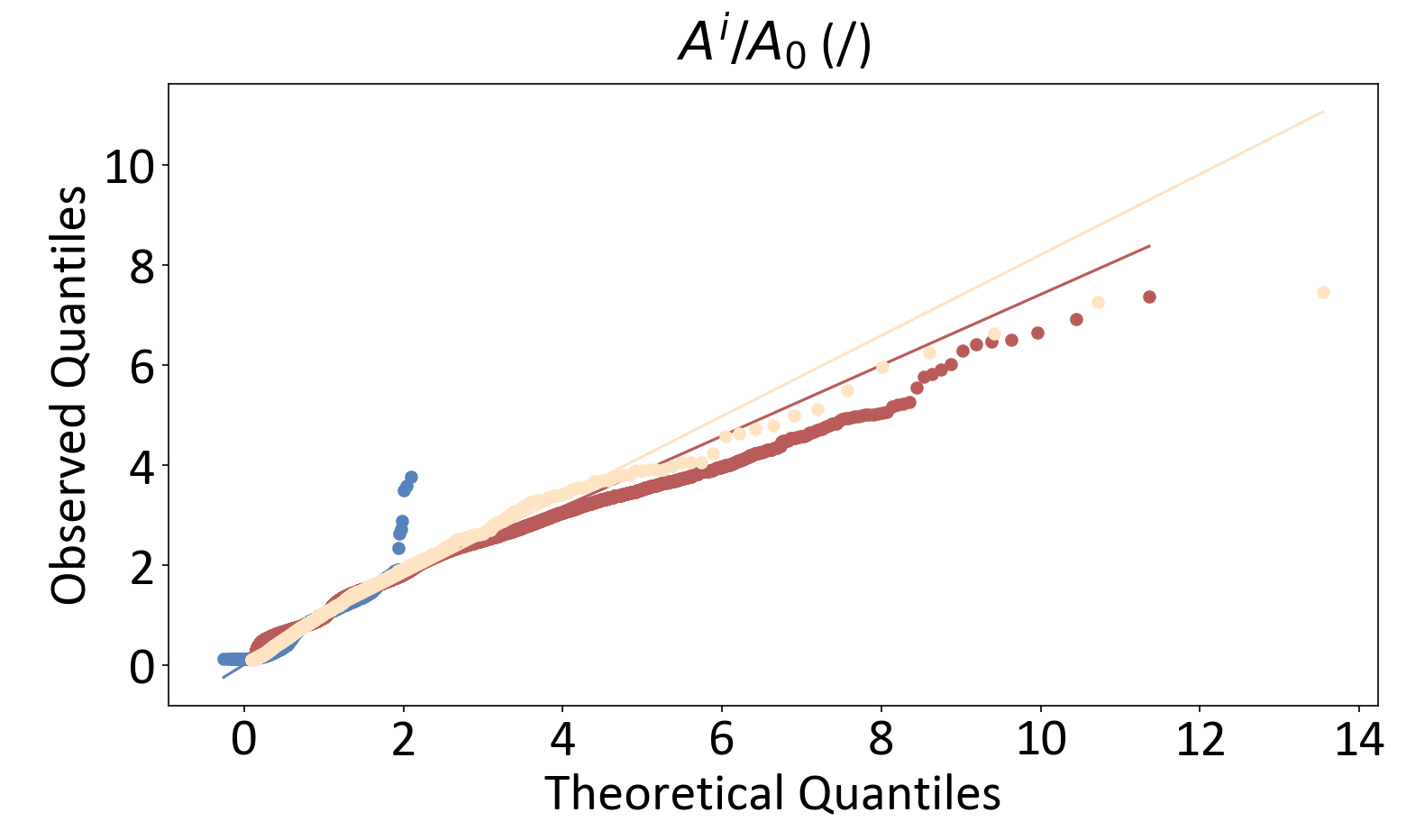}};
    \node (schematic) at (-2.5,0.7) {\includegraphics[width=0.05\textwidth]{network_stiffness_disorder_frame.png}};
    \node (schematic) at (-1,1.4) {\includegraphics[width=0.05\textwidth]{network_top_disorder_frame.png}};
    \node (schematic) at (3.5,1.4) {\includegraphics[width=0.05\textwidth]{network_top_disorder_2_frame.png}};
    \node[] at (-2.5,1.4) {\small{\textit{normal}}};
    \node[] at (-1.,2.1) {\small{\textit{lognormal}}};
    \node[] at (3.4,0.7) {\small{\textit{exponential}}};
\end{tikzpicture}
\caption{Relaxed pore size distribution of different network classes. Straight lines depict a comparison to distributions as noted for each case.}
\label{fig:quantiles_porosity}
\end{figure}

We finally quantify the resultant pore size distribution as a further microscopic network measure.
While the distribution of edge lengths has a direct influence on force propagation and the onset of damage, the relaxed pore size distribution relates to diffusion and the onset of phase separation within networks \cite{Fatin:2004,Ronceray:2022}. 
Figure \ref{fig:pores} illustrates relaxed pore size distributions across different disorder classes. 
In contrast to relaxed length distributions, pore sizes for networks displaying stiffness- and topological disorder display more complex distributions.
As illustrated in Figure \ref{fig:quantiles_porosity}, pore sizes in topologically disordered networks span the same range for disorder classes III and IV. However, deviations from lognormal and exponential distributions are observed in higher tail regimes.

\pagebreak
\newpage

\twocolumngrid

\section{Conclusion}

Macroscopically measurable mechanical properties are inherently determined by microstructural features. 
In the realm of flexible networks, the complexity of network structures encountered within the realm of soft matter constitutes a challenging modeling task.
Here, we take the reverse route of investigating the effect of generic network properties (length-, stiffness- and topological disorder) on the resultant network structure.
The chosen generic types of network disorder resemble different factors at play during polymerization (such as controlled homogeneity in crosslinking for purely length disordered networks).
We find that these different types of disorder in elastic networks lead to distinct relaxation characteristics:
At the microscopic scale, elastic networks displaying length- and stiffness disorder (disorder class II) follow a gaussian length distribution in their relaxed configuration.
This gaussian distribution is approached gradually with increasing stiffness disorder.
In stark contrast, topological defects instantly shift relaxation characteristics.
Topological defects induced in a regular network (disorder class III) result in lognormal relaxed length distributions.
Inherent topological defects stemming from a nearest neighbour search during network generation (disorder class IV) lead to an exponential relaxed length distribution, with tails leading up to large quantile ranges.
In practice, this corresponds to polymer networks with long tails in end-to-end distance, which include highly stretched chains that are expected to impact both the elastic and inelastic material response \cite{Tong:2018}.
Similarly, resultant pore sizes follow a gaussian distribution for networks displaying length- and stiffness disorder, whereas topological defects result in distributions that are more aligned with lognormally/exponentially distributed pore sizes.
These findings may be exploited to utilize more easily accessible experimental measures as a tool for inferring further details on network microstructure.
A lognormal distribution of local elastic stiffness as found in in biological tissues \cite{Millet:2021} for example hints at an underlying inhomogeneous topology, which may be harder to quantify.
At the macroscopic scale, we find that inducing topological disorder reduces the relaxed shear modulus, whereas increasing length disorder within the system increases the relaxed macroscopic shear modulus.
This behavior seems to be primarily governed by the inherent eigenstrain in fully flexible networks, with each network member being modeled as a coarse-grained freely jointed chain of restlength zero.

Several ways of extending this work could aid in completing the picture while remaining on a fully generic level:
The extension to semi-flexible networks would add further insights into possible changes in relaxation characteristics for networks which also exhibit bending rigidity.
In addition, accounting for inelastic networks allowing for failure would be an important extension in order to eliminate possible artifacts in mechanical properties due to high strain tail regimes.
Investigations of network topologies below the percolation threshold could furthermore shed light onto potential variations in relaxation behavior with increasing degrees of freedom.

\section{Acknowledgements}
SH gratefully acknowledges funding via the SNF Ambizione Grant PZ00P2186041. We thankfully acknowledge the computational efforts of David Buehler, as well as Dr. Alexander Ehret, Dr. Antoine Sanner and Dr. Robert W. Style for many helpful discussions.

\bibliography{apssamp}

\pagebreak
\newpage
\section{Appendix}

\begin{table}[!h]
\begin{center}
\begin{tabular}{ c|c|c|c|c| } 
  & I & II & III & IV \\ 
 \hline
 $<\text{number of edges}>$ & 8485 & 8485 & 6364 & 6537 \\ 
 \hline
 $<l_0>/l_0^{tri}$ & 1.0423 & 1.0423 & 1.0427 & 0.8436 \\ 
 \hline
 $<l_1>/<l_0>$ & 0.9594 & 0.9485 & 0.8444 & 0.6448 \\ 
 \hline
 $<z>$ & 5.85 & 5.85 & 4.39 & 4.85 \\
\label{tab:stats}
\end{tabular}
\caption{Network statistics across different network classes: average number of edges, initial mean length (normalized by length within a regular triangular network), and relaxed mean length (normalized by initial mean length). Values for network class III are given with $25\%$ deleted edges.}
\end{center}
\end{table}

\begin{figure}[!h]
\begin{tikzpicture}
    \node (schematic) at (0,1) {\includegraphics[width=0.5\textwidth]{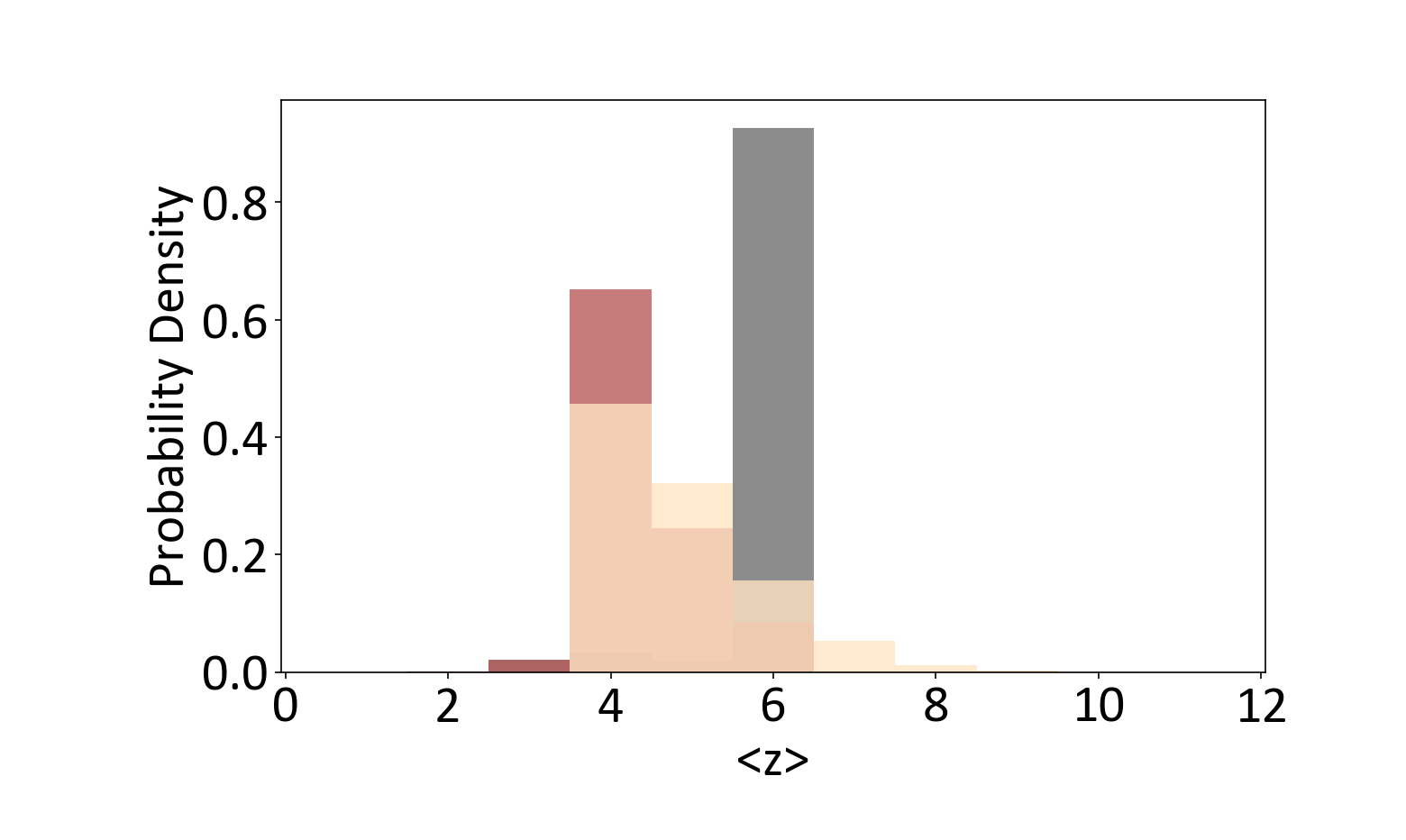}};
    \node (schematic) at (1.4,1.5) {\includegraphics[width=0.05\textwidth]{network_length_disorder_frame.png}};
    \node (schematic) at (2.5,1.5) {\includegraphics[width=0.05\textwidth]{network_stiffness_disorder_frame.png}};
    \node (schematic) at (-1.5,2.) {\includegraphics[width=0.05\textwidth]{network_top_disorder_frame.png}};
    \node (schematic) at (-1.5,0.5) {\includegraphics[width=0.05\textwidth]{network_top_disorder_2_frame.png}};
\end{tikzpicture}
\caption{Comparison of network degree distributions across network classes under investigation. Note that networks displaying length- and stiffness disorder (outlines in gray and blue, respectively) display the same distribution.}
\label{fig:network_degrees}
\end{figure}

\begin{figure}[!h]
\begin{tikzpicture}
-    \node (schematic) at (0,1) {\includegraphics[width=0.5\textwidth]{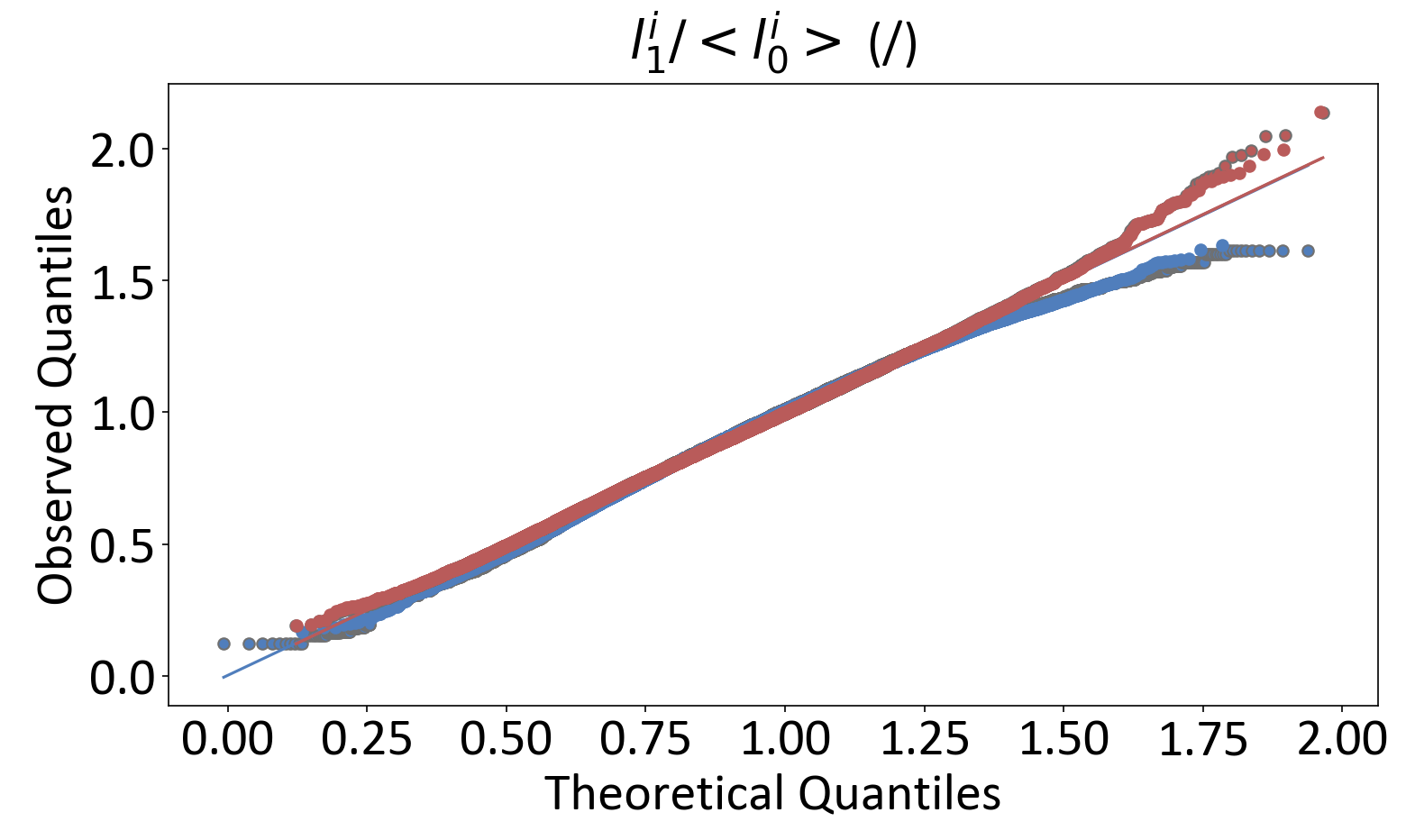}};
    \node[] at (3.6,0.9) {\small{\textit{normal}}};
    \node (schematic) at (3.6,1.5) {\includegraphics[width=0.05\textwidth]{network_stiffness_disorder_frame.png}};
    \node[] at (1.1,1.9) {\small{\textit{lognormal}}};
    \node (schematic) at (1.1,2.5) {\includegraphics[width=0.05\textwidth]{network_top_disorder_frame.png}};
\end{tikzpicture}
\caption{Relaxed length distribution for networks with stiffness- and topological disorder, comparing linear (no outline) and nonlinear constitutive behavior (gray outline).}
\label{fig:quantiles_linear_nonlinear}
\end{figure}

\begin{figure}[!h]
\begin{tikzpicture}
    \node (schematic) at (0,1) {\includegraphics[width=0.5\textwidth]{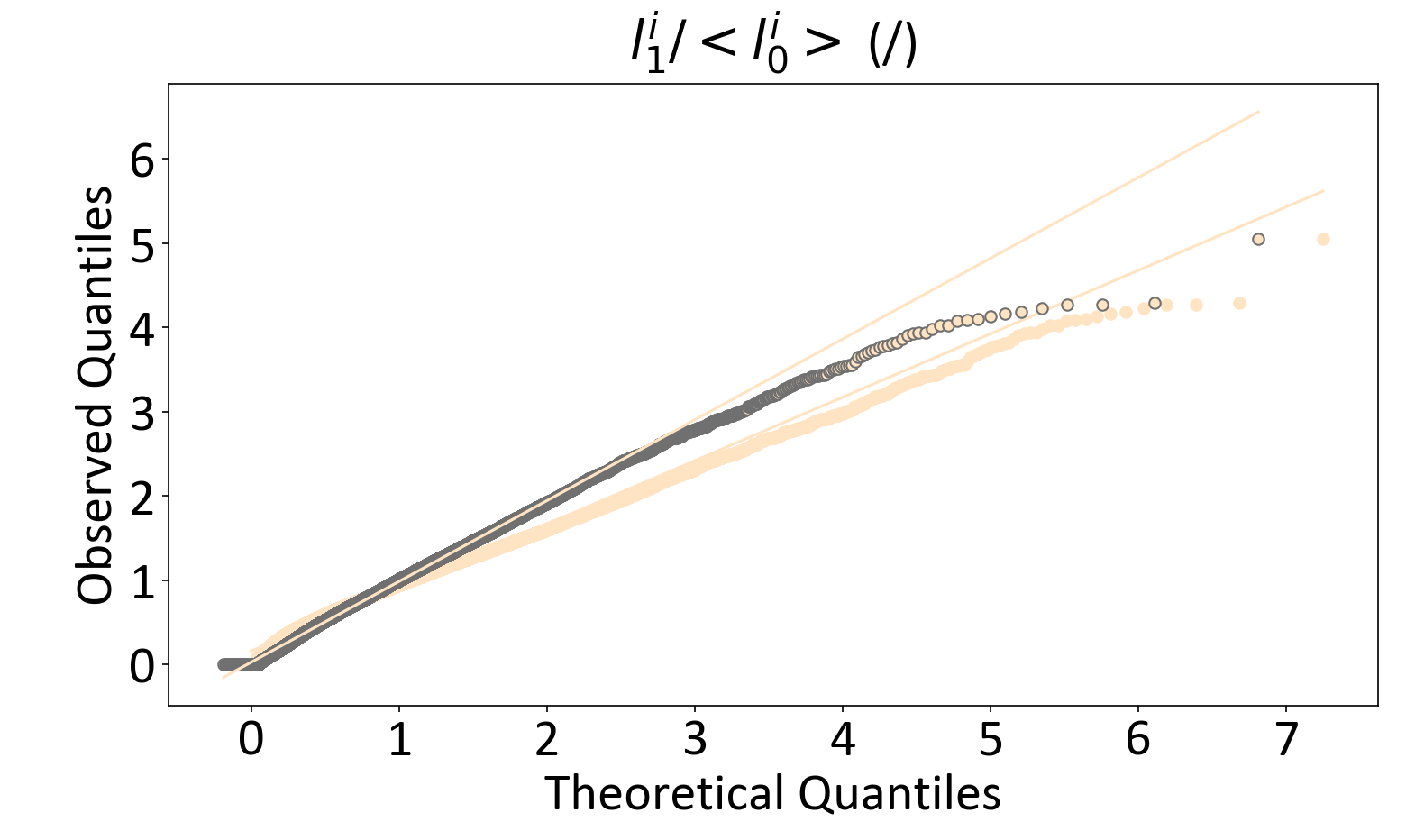}};
    \node[] at (3.4,0.55) {\small{\textit{exponential}}};
    \node (schematic) at (3.6,1.2) {\includegraphics[width=0.05\textwidth]{network_top_disorder_2_frame.png}};
    \node[] at (0.55,1.85) {\small{\textit{lognormal}}};
    \node (schematic) at (0.6,2.5) {\includegraphics[width=0.05\textwidth]{network_top_disorder_2_frame.png}};
\end{tikzpicture}
\caption{Relaxed length distribution for networks with topological disorder II, compared to exponentially distributed data (no outline) and lognormally distributed data (gray outline).}
\label{fig:quantiles_lognorm_exp}
\end{figure}

\begin{figure}[!h]
\begin{tikzpicture}
    \node (schematic) at (0,1) {\includegraphics[width=0.5\textwidth]{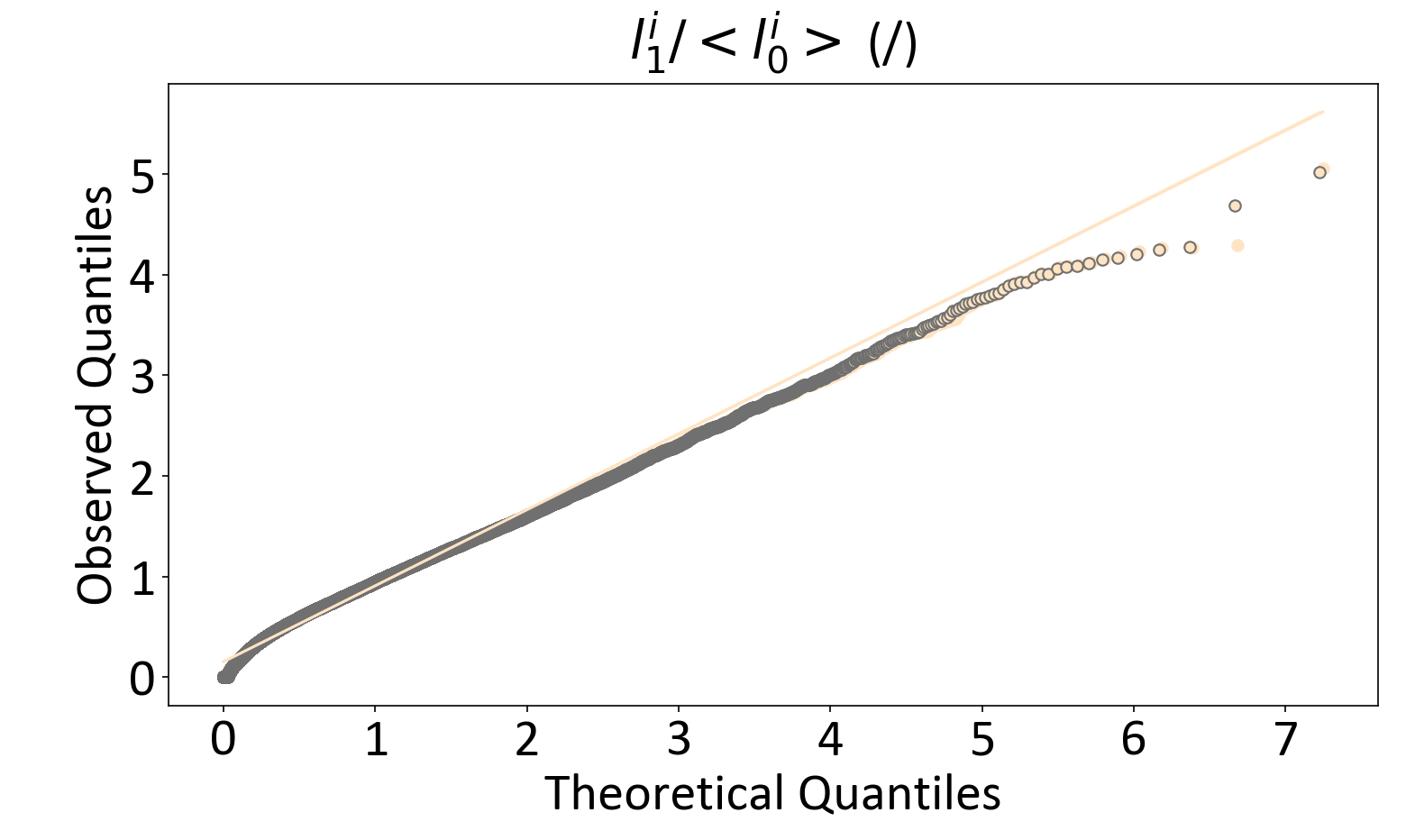}};
    \node[] at (3.,0.55) {\small{\textit{exponential}}};
    \node (schematic) at (3.2,1.2) {\includegraphics[width=0.05\textwidth]{network_top_disorder_2_frame.png}};
\end{tikzpicture}
\caption{Relaxed length distribution for networks with topological disorder II, comparing linear (no outline) and nonlinear constitutive behavior (gray outline).}
\label{fig:quantiles_linear_nonlinear_top}
\end{figure}

\end{document}